\documentclass[fleqn,usenatbib]{mnras}

\usepackage{newtxtext,newtxmath}

\usepackage[T1]{fontenc}
\usepackage{ae,aecompl}

\usepackage{graphicx}	
\usepackage{amsmath}	
\usepackage{amssymb}	

\usepackage[T1]{fontenc}
\usepackage{times}
\usepackage{url}
\usepackage{hyperref}
\usepackage{graphicx}
\usepackage{graphics} 
\usepackage{epsfig}   
\usepackage{microtype}
\usepackage{threeparttable} 
\usepackage{pdflscape}	

\def\apj{{ApJ}}                 
\def\apjl{{ApJ}}

\def\apss{ {Ap\&SS}}             
\def\aap{ {A\&A}}

\def\mnras{ {MNRAS}}

\def\nat{ {Nature}}

\newcommand{\cxo}{\textit{Chandra}}

\newcommand{\spi}{\textit{Spitzer}}

\newcommand{\be}{\begin{equation}}
\newcommand{\ee}{\end{equation}}

\newcommand{\ledd}{$L_{\rm Edd}$}

\newcommand{\gtsima}{$\; \buildrel > \over \sim \;$}
\newcommand{\ltsima}{$\; \buildrel < \over \sim \;$}
\newcommand{\prosima}{$\; \buildrel \propto \over \sim \;$}
\newcommand{\gsim}{\lower.5ex\hbox{\gtsima}}
\newcommand{\lsim}{\lower.5ex\hbox{\ltsima}}
\newcommand{\simgt}{\lower.5ex\hbox{\gtsima}}
\newcommand{\simlt}{\lower.5ex\hbox{\ltsima}}
\newcommand{\simpr}{\lower.5ex\hbox{\prosima}}

\newcommand{\lx}{$L_{\rm X}$}

\title[ALMA A0620--00]{ALMA observations of A0620--00: fresh clues on the nature of quiescent black hole X-ray binary jets}
\author[E. Gallo et al.]{Elena Gallo$^{1}$\thanks{E-mail: egallo@umich.edu}, 
Richard Teague$^{1}$
Richard M. Plotkin$^{2,3}$, 
James C. A. Miller-Jones$^{2}$, 
\newauthor
David M. Russell$^{4}$,
Tolga Din{\c c}er$^{5}$, 
Charles Bailyn$^{5}$, 
Thomas J. Maccarone$^{6}$,
\newauthor
Sera Markoff$^{7}$, 
Rob P. Fender$^{8}$\\ 
$^{1}$ Department of Astronomy, University of Michigan, 1085 S University, Ann Arbor, MI 48109, USA\\
$^{2}$ International Centre for Radio Astronomy Research -- Curtin University, GPO Box U1987, Perth, WA 6845, Australia\\
$^3$Department of Physics, University of Nevada, Reno, Nevada 89557, USA\\
$^{4}$ New York University Abu Dhabi, PO Box 129188, Abu Dhabi, UAE \\
$^{5}$ Yale University, Department of Astronomy, P.O. Box 208101, New Haven, CT 06520, USA\\
$^{6}$ Department of Physics and Astronomy, Texas Tech University, Box 41051, Lubbock, TX 79409, USA\\
$^{7}$ Anton Pannekoek Institute \& Center for Gravitation and Astroparticle Physics, University of Amsterdam, Science Park 904, 1098 XH Amsterdam, NL\\
$^{8}$ Department of Physics, University of Oxford, Denys Wilkinson Bldg, Keble Road, Oxford OX1 3RH, UK
}

\date{Accepted XXX. Received YYY; in original form ZZZ}

\begin{document}
\label{firstpage}
\pagerange{\pageref{firstpage}--\pageref{lastpage}}
\maketitle
\begin{abstract}
We report on ALMA continuum observations of the black hole X-ray binary A0620--00, at an X-ray luminosity nine orders of magnitude sub-Eddington. The system was significantly detected at 98 GHz (at $44 \pm 7~\mu{\rm Jy}$) and only marginally at 233 GHz ($20 \pm 8~\mu{\rm Jy}$), about 40 days later. These results suggest either an optically thin sub-mm synchrotron spectrum, or highly variable sub-mm jet emission on month timescales. Although the latter appears more likely, we note that, at the time of the ALMA observations, A0620--00 was in a somewhat less active optical-IR state than during all published multi-wavelength campaigns when a flat-spectrum, partially self-absorbed jet has been suggested to extend from the radio to the mid-IR regime. Either interpretation is viable in the context of an internal shock model, where the jet's spectral shape and variability are set by the power density spectrum of the shells' Lorentz factor fluctuations. While strictly simultaneous radio-mm-IR observations are necessary to draw definitive conclusions for A0620--00, the data presented here, in combination with recent radio and sub-mm results from higher luminosity systems, demonstrate that jets from black hole X-ray binaries exhibit a high level of variability -- either in flux density or intrinsic spectral shape, or both -- across a wide spectrum of Eddington ratios. This is not in contrast with expectations from an internal shock model, where lower jet power systems can be expected to exhibit larger fractional variability owing to an overall decrease in synchrotron absorption.
\end{abstract}






\section{Introduction}

The Atacama Large Millimeter Array (ALMA) offers the opportunity for the first time to characterize the properties of low-luminosity Galactic black hole X-ray binaries in the mm and sub-mm band. This window remains largely unexplored, and can prove to be crucial for our understanding of the interplay between accretion and the production of outflows at very low accretion rates.
Especially when it comes to winds and relativistic jets, our phenomenological knowledge (\citealt*{fender09}, and references therein) is largely based on 
observations taken during the bright phases of the hard X-ray state, i.e., at X-ray luminosity levels between about one thousandth to ten per cent of the Eddington luminosity (\ledd; \citealt*{plotkin13}). In this regime, the X-ray emission is ascribed to non-thermal radiation from a hot electron/positron population that enshrouds the inner accretion flow, either via Comptonization or synchrotron self-Compton, or a combination of the two. This is ubiquitously associated with persistent radio emission with a flat or slightly inverted spectrum extending up to IR/optical frequencies, where the outer disc and/or the donor star dominate the emission. The combination of the observed radio spectral indices, polarization levels and high brightness temperatures point to synchrotron radiation from a relativistic, collimated outflow, whereby the radio-emitting plasma becomes progressively more transparent at lower frequencies as it propagates toward larger distances from the base \citep*{bk79,kaiser06}. Arguably the best example of such a persistent, hard state, ``steady" jet is the high-mass black hole X-ray binary Cygnus X-1, whose persistent, flat (radio-mm) spectrum counterpart \citep{fender00} has been resolved into a compact core plus a variable, highly collimated jet with very long baseline interferometry \citep{stirling01}. In order to attain a steady, flat spectrum, the relativistic plasma within the jet must be constantly re-energized as it propagates downstream, so as to compensate for adiabatic energy losses. In analogy to both gamma-ray bursts and blazars \citep{spada}, repeated collisions between multiple plasma shells of different velocities/Lorentz factors have been proposed as a viable and efficient mechanism for providing the plasma with a constant energy replenishment \citep*{jamil,malzac2013}. Within the framework of this internal shock model, so long as the energy dissipation occurs at a (nearly) constant rate downstream in the jet, the net, \textit{time-averaged} result is a (nearly) constant flux density over a large frequency domain (i.e., up to the break frequency where the emission becomes optically thin), even though the physical nature of the so-called steady jet is intrinsically variable. Indeed, the Cygnus X-1 radio flux density displays 20--30 per cent level variability on time-scales of hours to months (\citealt{teta19cyg}, and references therein). 

Moving towards lower luminosities, it is generally believed that the jet properties are fundamentally unaltered, in that the flat radio spectra seem to be retained at Eddington ratios below $\simlt 10^{-6}-10^{-5}$, i.e. in the so-called quiescent regime. In a recent, comprehensive investigation of the radio variability properties of the black hole X-ray binary V404~Cygni in quiescence (at \lx$\simeq 3-4\times 10^{-5}$\ledd~\citealt*{bernardini14}), spanning over a quarter of a century, \cite{plotkin19} find evidence for relatively large deviations from a flat spectrum (with an average value of $\alpha=0.02\pm 0.65$, where $F_{\nu}\propto \nu^{\alpha}$). Moreover, 0.3--0.4 dex variability is found to be common on every observable timescale, from minutes up to decades (it should be noted that, owing to the sub-mJy flux densities involved, it is extremely challenging to spatially resolve V404 Cygni's quiescent radio counterpart into multiple, milliarcsec-scale beams with current instrumentation \citealt{millerjones08}). 
Within the internal shock model scenario, this behaviour can be interpreted as arising from a shock with higher-than-average amplitude propagating through an otherwise ``steady" (i.e., nearly constant dissipation throughout the) jet. Multi-wavelength investigations suggest that V404 Cygni's $\sim$flat synchrotron spectrum extends into the near-IR regime, where it becomes optically thin \citep{gallo07,hynes09,russellbreaks,plotkin15,plotkin17a}.

Our knowledge of jet behaviour is comparatively limited at extremely sub-Eddington ratios; in terms of radio properties, most of it hinges on radio detections of three nearby ($<3$ kpc) systems at \lx$\simlt 10^{-8}$ \ledd:
A0620--00 \citep{gallo06,gallo07,dincer18}, XTE J1118+480 \citep{gallo14,plotkin15}, and MWC~656 (\citealt*{dzib}; \citealt{ribo17}; see \citealt{millerjones11} for a compilation of upper limits). For all three systems, the measured radio flux densities are again indicative of synchrotron emission from a relativistic jet, as gyro-synchrotron emission from either donor star is estimated to be negligible. Moreover, the luminosities are consistent (within the large errors) with the extrapolation of the empirical, non-linear radio:X-ray luminosity correlation established for higher Eddington ratio systems (\citealt{corbel03}; \citealt*{corbel08}; \citealt{corbel13}; \citealt*{gfp03,gallo12}; \citealt{gallo14}). This has contributed significantly to the notion of quiescence as a dialed-down version of the more luminous hard state, whereby the jet power decreases along with the accretion flow's radiative efficiency.  A critical, open question, however, is whether, at such low Eddington ratios, the jet remains optically thick all the way to IR wavelengths. If so, and owing to the non-linear radio:X-ray luminosity scaling, then the jet -- rather than the accretion flow -- may dominate the power output from quiescent systems (\citealt*{fender03}; \citealt{gallo05a}; \citealt*{kording06a}; \citealt{russell10,coriat11,polko13,markoff15}).

Determining the full spectral extent of the radio-emitting jet in hard state and quiescent black hole X-ray binaries essentially translates into a careful modelling of the system's optical and IR emission. This aims to ascertain the presence of any excess IR emission with respect to the tail of the donor star's blackbody spectrum, and to determine its physical nature. In addition to synchrotron radiation from a jet, excess IR emission can be ascribed to a variety of mechanisms, including synchrotron radiation from a hot, radiatively inefficient inflow, thermal emission from the accretion stream-outer flow impact point, thermal emission from the outer accretion disc and/or thermal emission from a cold, dusty disc of circumbinary material (see \S~\ref{sec:disc} for details and references). 

At \lx$\simeq 10^{-9}$ \ledd, and $1.06\pm0.12$ kpc distance \citep{cantrell10}, the quiescent black hole X-ray binary A0620--00 provides us with a testbed for competing models through a direct measurement of its mm spectrum. In this Letter, we report on dual band ALMA observations of A0620--00, obtained in November and December 2016. In concert with closely-spaced radio and IR-optical observations, these data enable us to place new constraints on the jet properties in this prototypical system. 

\section{Observations and data reduction}

\subsection{ALMA}

Continuum observations in Band 3 at 98~GHz and Band 6 at 233~GHz were taken as part of project 2016.1.00773.S (P.I. Gallo). The Band 3 data were taken on the 12th and 14th November 2016 with 46 antennas, with baselines spanning the range 15\,m to 1039\,m, and for a total of 1.1 hr of integration. The Band 6 data were taken on the 20th and 21st December 2016 with 48 antennas, baselines ranging between 15\,m and 491\,m, for 2.15 hr. All observations were performed with good weather conditions with median precipitable water vapor of 0.36~mm for the Band 3 observations and 1.29~mm for Band 6.

\begin{figure}
\hspace{-0.2cm}
    \includegraphics[scale=0.6]{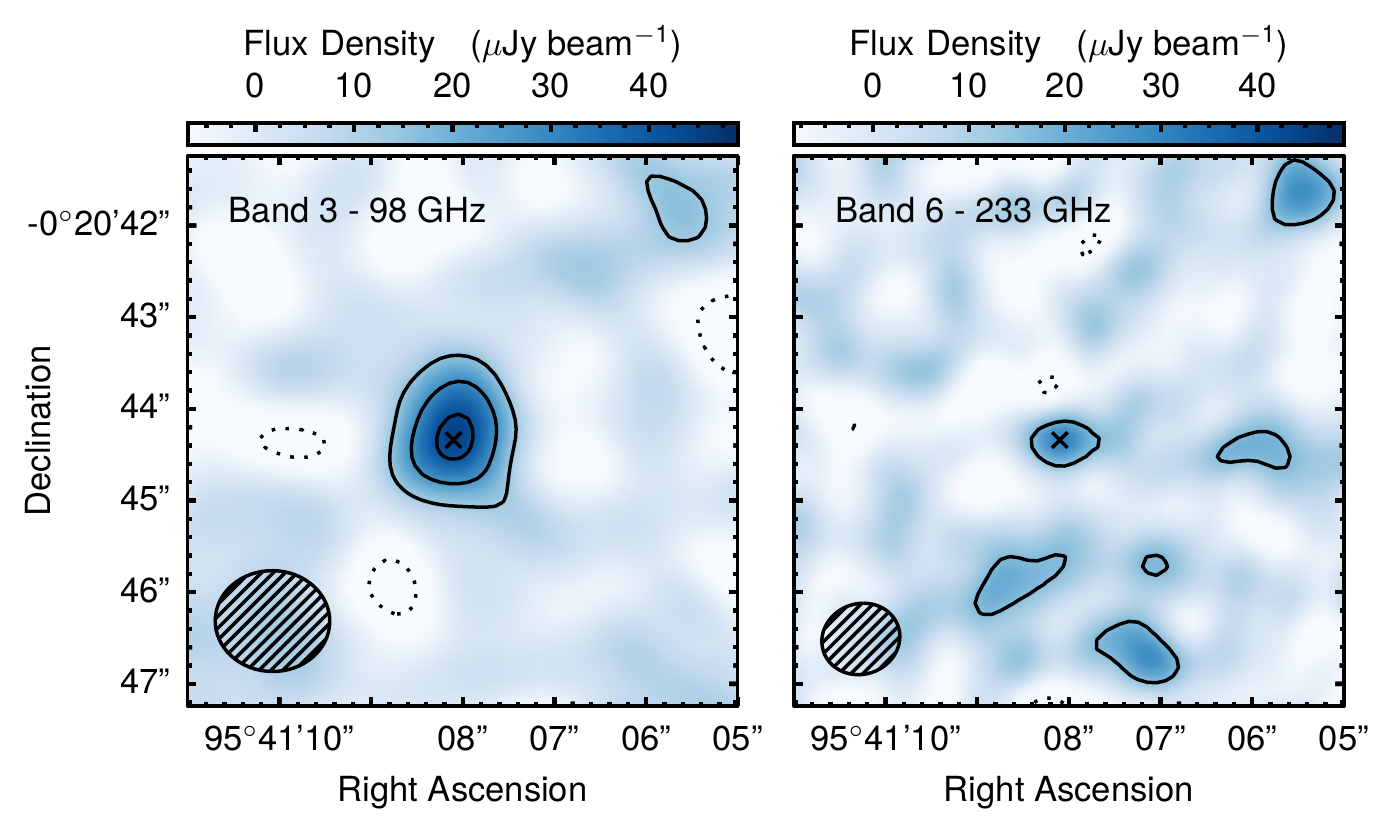}
    \caption{ALMA continuum images at 98~GHz (band 3), left, and 233~GHz (band 6), right. The contours are in steps of $2\sigma$ where $\sigma = 7.9~\mu{\rm Jy~beam}^{-1}$ for Band 3 and $8.2~\mu{\rm Jy~beam}^{-1}$ for Band 6, with dotted contours being negative. The source center is shown by the black cross, and the synthesized beams are shown in the lower left corner of each panel.}
    \label{fig:ALMA_continuum}
\end{figure}

The data were calibrated using $\texttt{CASA}$ \citep[v5.1;][]{McMullin_ea_2007} using the pipeline script provided by ALMA staff. For both sets of observations the quasars J0750+1231 and J0641-0320 were used for flux and phase calibration, respectively. After the initial calibration, there was insufficient signal to perform self-calibration.
The data were imaged using natural weighting to maximize sensitivity, achieving a beam size of $1.25\arcsec \times 1.10\arcsec$ with a position angle of $89.0\degr$ for Band 3, and a root mean square (rms) noise of $6.9~\mu{\rm Jy~beam}^{-1}$. For the Band 6 data the synthesized beam is $0.86\arcsec \times 0.77\arcsec$ at a position angle of $-71.3\degr$, with an rms noise of $8.2~\mu{\rm Jy~beam}^{-1}$. The resulting images are shown in Fig.~\ref{fig:ALMA_continuum}. A0620--00 is clearly detected in the Band 3 continuum, while only marginally significant emission is found for Band 6.
In order to check whether phase decorrelation could have reduced the detection significance in Band 6, we reran the calibration pipeline, flagging three scans on the phase calibrator (scans 8, 14 and 20) prior to the gain calibration, and treating them as a second target source, interpolating the complex gains from neighbouring calibrator scans.  Imaging the data from these three scans gave a peak flux density within 5 per cent of that measured from the self-calibrated data on the same source.  This confirms that phase decorrelation as a function of time was $<$5 per cent during the Band 6 observations. While we cannot explicitly test for decorrelation as a function of position on the sky, the time stability and Band 6 weather conditions give us a degree of confidence that the (5.6 degree) positional shift between the phase calibrator and A0620--00 was unlikely to be responsible for decorrelation as a function of position.

The integrated flux densities were measured by fitting a 2D Gaussian with the same properties as the synthesized beam to the continuum images, allowing the total integrated flux density and source position to vary. Posterior distributions were sampled using an affine invariant Markov chain Monte Carlo ensemble sampler, \texttt{emcee} \citep{Foreman-Mackey_ea_2013}. In Band 3 the best-fitting integrated flux was $43.6 \pm 0.8~\mu{\rm Jy}$ and in Band 6 it was $20 \pm 1~\mu{\rm Jy}$, where the 16th to 84th percentiles of the posterior distribution (which was symmetric about the median) were used to quantify the uncertainty in the Gaussian's amplitude. 
Absolute flux calibration adds a systematic uncertainties of 5 per cent for Band 3 and 10 per cent for Band 6; overall, the measurement errors are dominated by rms noise uncertainties, yielding a robust detection in Band 3, at $44 \pm 7~\mu{\rm Jy}$, and a marginally significant detection in band 6, at $20 \pm 8~\mu{\rm Jy}$.

\subsection{VLA}
We obtained 13 observations with the Karl G.\ Jansky Very Large Array (VLA) between the 11th September 2017 and 22nd January 2018 (program 17B-233; PI: Plotkin).  Each observation lasted between 30 and 90 min, for a total of 666 min integration time on source across all 13 observations.  The data  were taken in continuum mode at X-band (8--12 GHz) using 4 GHz total bandwidth (over 2$\times$2 GHz basebands centred at 9.0 and 10.65 GHz).   Data were processed using {\sc CASA} v5.1.1 \citep{McMullin_ea_2007}.  We used calibrated data products provided by NRAO from the VLA CASA Calibration pipeline v5.1.2.  For these calibrations, for 11/13 observations, 3C~48 was used as the primary calibrator to perform delay calibration, to find complex bandpass solutions, and to set the flux density scale (3C~286 was used for the other two observations).  All 13 observations used J0643+0857 as a nearby secondary calibrator to derive the complex gain solutions.  
Imaging was performed using {\sc clean} in {\sc CASA}.  We stacked all 13 images, but imaged each 2 GHz base-band separately to obtain spectral information.  We used natural weighting to maximise the sensitivity, and we accounted for the frequency dependence over the large fractional bandwidth using two Taylor terms ({\sc nterms=2}).  We measured flux densities in the stacked images of each 2 GHz baseband using {\sc imfit}, forcing a point source (fixed to the size of the synthesized beam) at the known location of A0620--00.  We measured flux densities of $12.9\pm1.5$ and $14.2 \pm 1.8$ $\mu$Jy bm$^{-1}$ at 8.9 and 10.8 GHz, respectively.  The quoted error bars correspond to rms errors for each image, conservatively inflated by 10 per cent to account for potential systematic uncertainties in the flux density scale derived from 3C~48, which started a large flare around 2018 January (towards the end of our observations).  Radio variability is estimated to be well within a factor of 2 during this monitoring campaign (Plotkin et al., in prep.).

\subsection{SMARTS}
A0620--00 is routinely monitored at optical and near-IR (NIR) frequencies with the dual-channel imager ANDICAM on the SMARTS 1.3~m telescope, at the Cerro Tololo Inter-American Observatory. Data presented here were reduced following standard procedures described in \cite{dincer18}; point-spread function photometry was first performed to measure instrumental magnitudes for A0620--00 and several nearby field stars, and then converted to the standard photometric system through differential photometry, with absolute calibration via optical primary standards on clear nights and the Two Micron All-Sky Survey catalog. 

\begin{figure}
\includegraphics[width=0.5\textwidth]{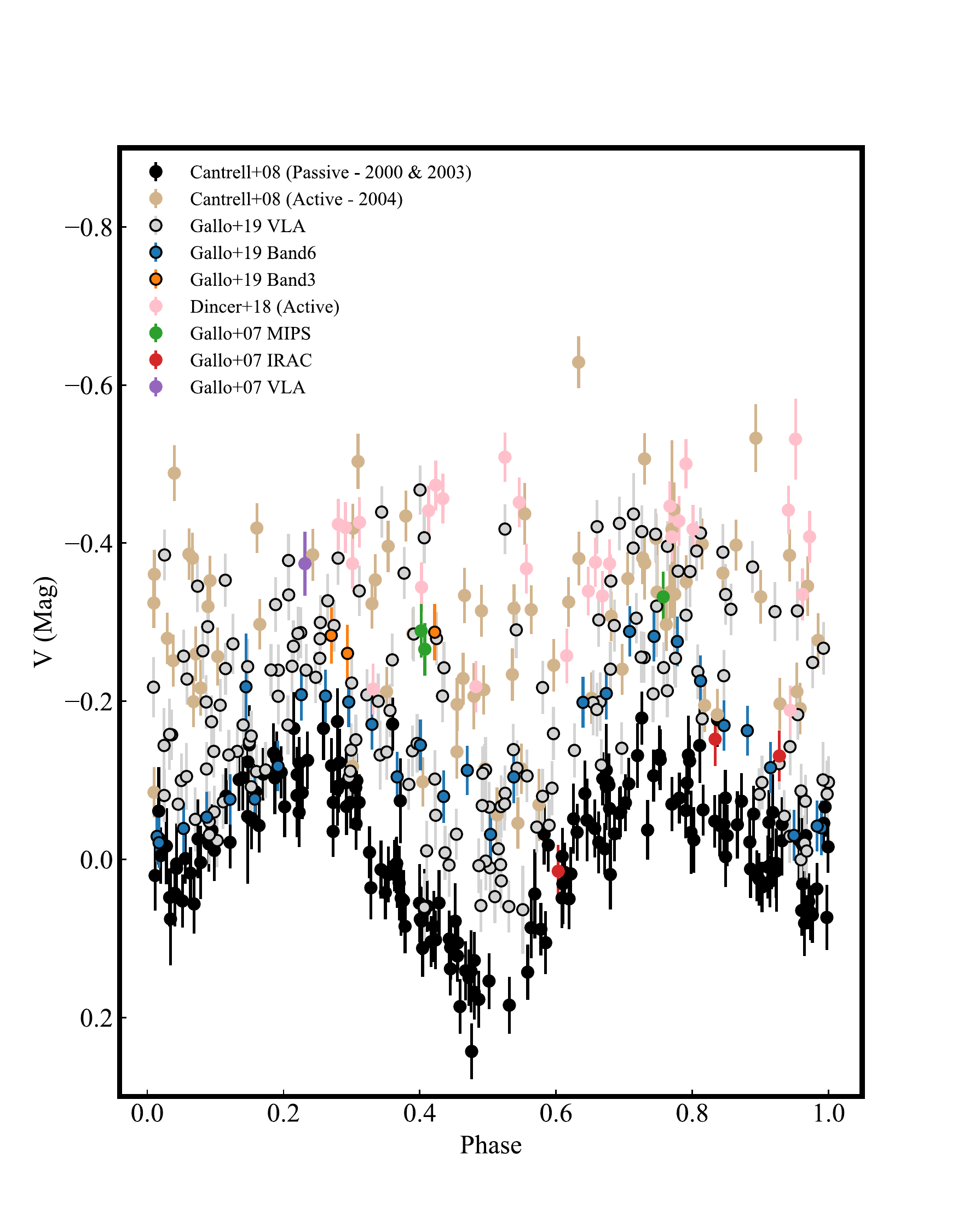}
\caption{A0620--00's orbital-phased SMARTS V-band magnitudes, calculated using the updated ephemeris by \protect\cite{cherepashchuk19}. Nearly simultaneous epochs with other multi-wavelength measurements quoted in the body of this paper are highlighted with different colors, and compared against the passive state ``envelope" defined by Cantrell et al. (2008), in black, vs. active state data from  Cantrell et al., as well as \protect\cite{dincer18}, in sand and light pink. Blue and orange points, respectively, are simultaneous within $\pm$2 days with the 2016 Band 6 and Band 3 ALMA data presented here. While they are formally inconsistent with a pure passive state, the Band 3 data (which, unlike Band 6, resulted in a significant detection) may correspond to a slightly higher activity level. \label{fig:pa}}
\end{figure}

For calculating the intrinsic source fluxes, we used the zero points given in \cite*{bessell98}, and the color excess $E(B-V)=0.30\pm 0.05$ (Cantrell et al. 2010), which was converted to total extinction values ${A}_{{\rm{B}}}=1.23$, ${A}_{{\rm{V}}}\,=0.93$, ${A}_{{\rm{I}}}=0.47$, ${A}_{{\rm{J}}}=0.26$, ${A}_{{\rm{H}}}=0.16$, and ${A}_{{\rm{K}}}=0.11$. Shown in Figure \ref{fig:pa} are A0620--00's, phase-folded, SMARTS V-band magnitudes as measured over a variety of epochs close to the multi-wavelength campaigns discussed below. This is for the purpose of illustrating whether, at the time of a specific campaign, the source was in a passive vs. active quiescent state, as identified by Cantrell et al. (2008; see \S~3.1). Figure \ref{fig:sed} shows instead the average H-I-V and B-magnitudes as measured on December 20 and 21, 2016, i.e. simultaneous with the ALMA Band 6 data.

\section{Results and discussion}\label{sec:disc}

To interpret the ALMA observations presented here, it is useful to start by critically reviewing the relative complexity of what we refer to as quiescence, and the many ways the black hole X-ray binary A0620--00 has played a key role in our understanding of it. 
\subsection{A0620--00's jet: spectral extent and variability}
In a seminal work, \cite{cantrell08} analysed optical and NIR orbital-phase resolved data of A0620--00 from 1981 to 2007 (see also \citealt{cantrell10}). They first distinguished between two main quiescent states -- passive and active, plus loops state, in between the two -- each representing different levels of activity. Passive state light curves display the lowest level variability; in this regime, the ellipsoidal modulation of the donor star fully accounts for the observed optical-NIR variability. In the active state, the optical--NIR flux is 0.1--0.4 magnitudes brighter and variable, with prominent variations even on short timescales (down to seconds; e.g., \citealt{hynes03c}; \citealt*{zurita03}; \citealt*{neilsen08}).
A0620--00, which has been in quiescence since the decline from the 1975 outburst that led to its discovery \citep{elvis75}, can spend months or years in these different quiescent states, being in the passive state in 1997--2003, the active state in 2004--2007 and 2013, the passive state in 2015--2016, changing to the active state by 2016 November \citep{shugarov16,vangrunsven17,dincer18,cherepashchuk19}.  Figure \ref{fig:pa} illustrates the phase-folded, V-band variability of A0620--00 during (or close to) the different campaigns and observations discussed below, vis-a-vis the well-identified passive and active states, shown by the black and sand/pink circles, respectively. 

Several lines of evidence suggest that, {at least during the active state}, the spectral properties of the radio jet in A0620--00 resemble those of higher Eddington ratios systems, such as V404 Cygni, where a partially self-absorbed synchrotron spectrum extends all the way from radio up to mid-IR wavelengths \citep[e.g.][]{russellbreaks,russell13,teta15,maitra17}. Arguably the most compelling piece of evidence for this comes from broadband SED modelling. With an 8.5 GHz counterpart at 51$\pm 7 \mu$Jy, A0620--00 was the first truly quiescent black hole X-ray binary to be detected in the radio band, in 2005 \citep{gallo06}. Earlier observations, performed in 2003 with \spi\ MIPS had shown evidence for excess mid-IR emission with respect to the tail of the donor star at 8 $\mu$m \citep*{muno}. These authors {ascribed the mid-IR excess to thermal emission from a cold, circumbinary dust disc}, illuminated by the low-mass donor star (such discs may be formed as a result of mass outflow from the outer accretion disc; \citealt*{taamspruit}). Subsequently, \cite{gallo07} combined optical and X-ray observations performed in 2005 (simultaneously with the VLA) with the (non-simultaneous) \spi\ MIPS data (all in active state -- see Figure 2), and reinterpreted the mid-IR excess as arising from the extrapolation of a $\sim$flat-spectrum radio jet extending all the way to the \spi\ band (if true, this would imply that the radiative output of the jet in A0620--00 would be comparable to -- if not greater than, depending on the cooling break frequency -- that of the accretion flow).

Along the same lines, \cite{froning11} fitted the broadband SED of A0620--00, taken in 2010 March (likely during the active state, according to the same authors; no SMARTS data are available for this campaign), and including high-resolution optical and UV spectra from Keck and the \textit{Hubble Space Telescope}. They found that the non-stellar light had a peak at 0.3 $\mu$m that could be fitted by a black-body, likely from the hot spot/stream impact point onto the accretion disc from the donor star. They also reported on a UV upturn in flux and a red excess that can both be fitted well by a model in which a partially self-absorbed jet extends all the way to the mid-IR, with the pre-acceleration inner jet component dominating the UV upturn, and the post-acceleration synchrotron dominating the red excess towards the IR (simultaneous radio observations with the Australia Telescope Compact Array did not detect A0620--00 down to 42 $\mu$Jy at 5.5 GHz). \cite{froning11} inferred from their modelling that the mass accretion rate from the donor through the hot spot \citep*[which has also been detected in Doppler maps from optical emission lines;][]{gonzalez10} was five orders of magnitude higher than the accretion rate at the black hole inferred from the X-ray luminosity, implying either highly radiatively inefficient accretion, and/or that outflows expel almost all of the accreted mass, and/or the mass transfer rate from the outer disc into the inner hot region is very low as matter builds up in the disc.\\

Optical--IR variability also gives important clues to the nature of the IR emission in A0620--00, and other quiescent systems. The power density spectrum of the highly variable optical-IR component detected in the active quiescent state, when measured, is similar to that of an X-ray power density spectrum of a black hole X-ray binary in the hard state, in which the X-ray variability originates in the inner regions of the accretion flow. This ``flickering" component is stronger at longer wavelengths, with a fractional rms amplitude of $\sim 15$--24 per cent at optical wavelengths, rising to $\sim 42$ per cent at ($K$-band) NIR wavelengths \citep{dincer18}. The spectral index of the variable component has been inferred by various authors to be steeply red at optical--NIR wavelengths ($\alpha = -1.4$ to $-0.7$; \citealt{shahbaz04,cantrell10,dincer18}), and flat/slightly inverted at NIR ($\alpha \sim 0.3$ to 0.4; \citealt{cherepashchuk19}) and mid-IR ($\alpha = 0.2$ to 0.3 at 3.6--8.0 $\mu$m; \citealt{maitra11}) wavelengths. Noting that the extrapolation of this non-stellar mid-IR spectrum of A0620--00 down to GHz frequencies was consistent with the radio flux measured by the VLA in 2005, \cite{russellbreaks} concluded that the optically thick-to-thin jet break lies in the optical frequency range: $1.3\pm0.5\times 10^{14}$ Hz.  

Finally, \cite{russell16_pol} reported an excess of linear polarization of $\sim 1.25 \pm 0.28$ per cent in the NIR for A0620--00 (from observations taken in February 2013, again during the active state), with a position angle of the magnetic field vector that is consistent with being parallel with the inferred axis for the transient radio ejection associated with the source 1975 outburst \citep{kuulkers99}, providing further circumstantial evidence for the jet interpretation of the IR excess -- albeit alternative mechanisms, such as coronal synchrotron emission or a circumbinary disc whose dust grains align with the global magnetic field of the accretion disc may be able to yield similar polarization signatures. \\

In spite of multiple, indirect lines of evidence for a $\sim$flat radio-mid-IR jet spectrum in the active state of A0620--00 \citep[and also in Swift J1357.2--0933, in which the jet break has been found in the IR in quiescence,][]{plotkin16,russell18}, there are reasons to remain skeptical. It should be noted that, to first order, the jet break frequency is expected to scale inversely with the black hole mass \citep*{falcke95,heinz03}; as jet breaks are often observed in the GHz/sub-mm regime in active galactic nuclei, they are expected to occur in the IR-optical band for $10^{5-7}$ times lighter objects. At the same time, additional parameters are thought to affect the exact jet break frequency for a given system: e.g., the exact value is known to vary with the overall luminosity level as well as X-ray photon index (\citealt*{homan05}; \citealt{gandhi11,koli15,vincentelli18}), likely reflecting changes in the magnetic field energy density, particle density and/or mass loading at the jet base \citep{markoff10,russellbreaks,markoff15}. While an optical jet break for a system at nine orders of magnitude sub-Eddington remains somewhat challenging in the context of basic scale-invariant jet models, recent developments allow for more nuanced solutions. As an example, the 3 dex excursion in jet break frequency observed during a single state transition in the black hole X-ray binary MAXI~J1836--194 \citep{russellt14} has been successfully modelled through a semi-analytical treatment of the relativistic-magneto-hydrodynamic jet equations (\citealt*{polko14}; \citealt{ceccobello18}), where the jet break frequency is assumed to correspond to the location of the jet's ``modified fast point". Alternatively, it may be that an additional spectral component contributes some of the optical emission.

The last broadband SED investigation of A0620--00 prior to our ALMA observations was presented by \cite{dincer18}, who collected simultaneous radio, IR, optical and X-ray data in December 2013 (see pink points ``Dincer+18" in Figure 2; also in the active state). While they did not attempt an overall spectral fit, \cite{dincer18} were the first to report a highly inverted radio spectrum, between 5--22 GHz, with $\alpha={+0.70\pm 0.13}$.  Taken at face value, this spectrum is inconsistent with an extrapolation of the known mid-IR excess to radio frequencies; in the absence of mm data, however, it could either signal a highly peaked radio-mm spectrum, or simply trace high-level variability, or, the low frequency curvature of an otherwise $\sim$flat 50 $\mu$Jy spectrum above 20 GHz, (as seen, e.g., in V404 Cygni; \citealt{russellbreaks,plotkin19}).

\begin{figure}
\includegraphics[width=0.5\textwidth]{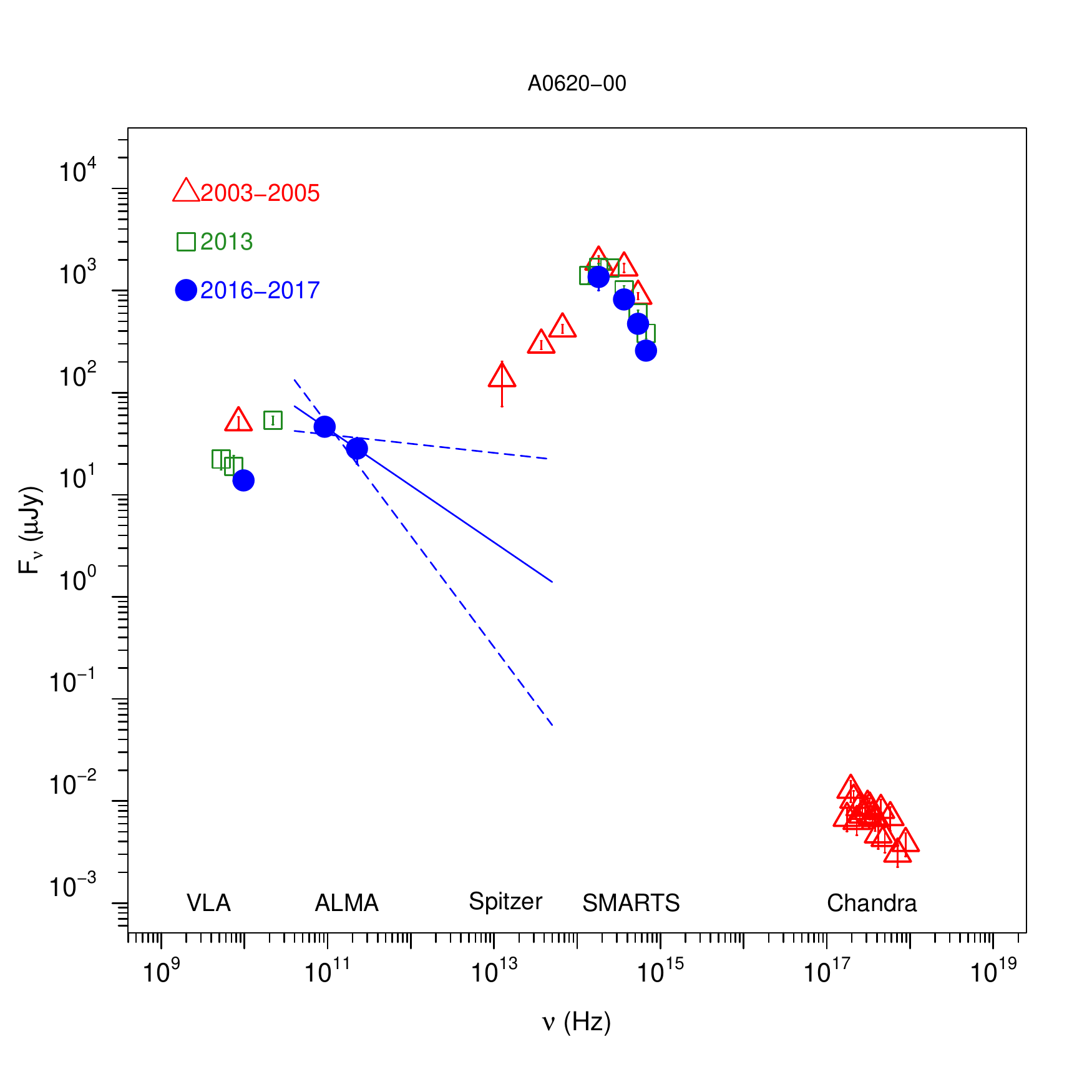}
\caption{Broadband SED of A0620--00 in quiescence, over different epochs. Red open triangles are based on data taken over 2003-2005 (VLA, \textit{Spizter}, SMARTS and \textit{Chandra}) and reported by Gallo et al. (2007); green open squares refer to the 2013 campaign (VLA and SMARTS) reported by (Din{\c c}er et al. 2018); blue filled circles represent the most recent data set, consisting of the 2017 VLA and 2016 ALMA observations presented in this work, combined with SMARTS data.\label{fig:sed}}
\end{figure}

Figure \ref{fig:sed} shows the broadband SED of A0620--00~for three of the multi-wavelength campaigns discussed above: red open triangles correspond to the 2003-2005 campaign reported by Gallo et al. (2007; VLA, SMARTS and \cxo), including the \spi~data from \cite{muno}; green open squares refer to the 2013 campaign (VLA and SMARTS) reported by \cite{dincer18}; blue filled circles represent the most recent data set, consisting of the new VLA, ALMA and SMARTS observations presented in this work.
The ALMA data alone could easily be interpreted as indicative of optically thin synchrotron emission, with $\alpha={-0.56\pm 0.47}$ (3$\sigma$ C.L.) over the 98--233 GHz frequency interval. However, as the Band 3 and 6 observations were taken on 12-14th November and 20th-21st December 2016, respectively, it is also possible -- and arguably more likely -- that the measured values indicate a factor $\simgt 2$ variability in flux density over $\simgt 1$-month timescale. This would be broadly consistent both with the measured variability at 5--8 GHz over a time-scale of several years (see again Figure \ref{fig:sed}), as well as with the much shorter term variability inferred for the jet of V404 Cygni at radio frequencies (albeit at 3 dex higher \ledd; \citealt{plotkin19}). 

That intrinsic variability may be at the origin of what we are observing is indirectly supported by the SMARTS monitoring, as illustrated in Figure \ref{fig:pa}, where data from within $\pm 2$ days of the 98 GHz (Band 3) and 233 GHz (Band 6) ALMA observations, respectively, are shown as orange and blue points. While both ALMA epochs appear inconsistent with a pure passive state  (i.e. with the black points), the 98 GHz data (which resulted in a significant detection) broadly overlap with active state data from previous campaigns (light pink and sand points), and, seem to correspond to a slightly higher activity level than the 233 GHz data. At the same time, it should be noted that, during both ALMA observations, A0620--00 was somewhat less active than during other published multi-wavelength campaigns discussed above, when a flat-spectrum, partially self-absorbed jet been suggested to extend from the radio to at least the mid-IR regime.

\subsection{On the variable nature of black hole X-ray binary jets}
While strictly simultaneous radio-mm-IR observations are necessary to draw definitive conclusions, the data presented here, when discussed in the context of previous multi-wavelength campaigns, suggest that A0620--00's jet has a highly variable nature. 
Whether they are best explained by a variable jet break location (with a $\simlt 100$ GHz break frequency, compared to previous claims in the mid-IR/optical), or significant intrinsic flux variability within an otherwise partially self-absorbed jet, the 2016 ALMA observations of A0620--00, in combination with recent radio results from V404~Cygni \citep{plotkin19} and Cygnus X-1 \citep{teta19cyg}, and optical--IR results from Swift J1357.2--0933 \citep{shahbaz13,russell18}, demonstrate that \textit{jets from black hole X-ray binaries exhibit a high level of variability across a wide spectrum of Eddington ratios.}

In the context of the internal shock model, A0620--00's behaviour may be linked to a systematic change in the amplitude and/or injection time-scale of the shells' Lorentz factor fluctuations. With respect to the latter, \cite{malzac2014} find that, whereas a $\sim$flat radio-mid-IR jet SED naturally arises from a flicker noise process, i.e., where the power spectral density of the Lorentz factor fluctuations is inversely proportional to the Fourier frequency ($P(f)\propto 1/f$), a peaked spectrum at GHz frequencies can result from a steep power spectral density, where $P(f) \propto 1/f^{\beta}$, with $\beta>1$. For steeper $\beta$ values, the Lorentz factor fluctuations have, on average, longer time-scales, meaning that most of the shell collisions, and hence energy dissipation, occur farther away along the jet (see figures 5e and 5e in \citealt{malzac2014}). This would correspond to a scenario where the jet break frequency indeed occurred below $\sim$ 100 GHz at the time of the ALMA observations. 
Alternatively -- and arguably more likely -- the ALMA data could signal $\simgt 100$ per cent flux variability over a month timescale. Going back to the internal shock model, a broad range of variability timescales can be induced if/when higher-than-average amplitude fluctuations propagate through an otherwise fainter, flat-spectrum jet which may or may not extend to the mid-IR (this has also been suggested to explain the optically thin radio spectral index that is sometimes observed in V404 Cygni; \citealt{plotkin19}). Incidentally, for a conical jet geometry, the fractional variability is bound to be progressively higher at higher frequencies, where the size of the emitting region becomes progressively smaller (in spite of the very limited diagnostics available in the mm and sub-mm windows, observations of V404 Cygni during the decay from its 2015 outbursts appear to confirm enhanced variability at sub-mm wavelengths, compared to cm; \citealt{teta19v404}).  \\

In the specific case of A0620--00, it is important to stress again that the ALMA observations were taken at a time when the system's optical-IR variability had not fully reached the active quiescent state, which previous investigations have claimed to be associated with a mid-IR jet.  It is interesting to recall that, in their seminal work, \cite{cantrell08} ascribe the transition from passive to loops and active states to the onset of clumpy accretion, where the loops are ``qualitatively consistent with expanding shells of gas, initially heated to be bluer than A0620--00's secondary, then cooling adiabatically as they redden significantly while remaining bright, then finally fading from view". The active state is described as more erratic, with large color as well as magnitude fluctuations that are seldom as faint as the minimum passive state data.  We speculate that the passive quiescent state may correspond to a regime where the jet emission is negligible at mid-IR frequencies -- possibly even suppressed entirely as a result of extremely low accretion rates. The sudden onset of (clumpy) accretion events would initiate the building up of a jet. Initially, the power density spectrum of the Lorentz factor fluctuations could be fairly steep ($\beta \simgt 1 $), possibly corresponding to a peaked sub-mm/mm synchrotron spectrum, and would progressively flatten as the system settles into the active quiescent state. If so, the jet itself may ultimately be responsible for \textit{defining} the active and passive quiescent states of A0620--00 as identified by Cantrell et al. More importantly, this behaviour should be common at extremely low accretion rates. Higher Eddington-ratios systems, where accretion takes place in a more sustained fashion, would instead be characterized (in A0620--00's jargon) by a \textit{persistent} active state, where the jet is likely responsible for the observed mid-IR excess at all times. \\

In closing, no firm conclusions can be drawn based upon on the dual band ALMA observations of the quiescent black hole X-ray binary A0620--00 presented here. The source was significantly detected at 98 GHz (Band 3) in mid November 2016, and only marginally so at 233 GHz (Band 6) some 40 days later. This can either be interpreted as the signature of a partially self-absorbed jet whose break frequency occurred somewhere below 98 GHz, or, more simply, in terms of intrinsic flux variability on $\sim$month timescales in the ALMA bandpass. Strictly simultaneous radio and mm observations of A0620--00, in concert with IR and optical, are necessary to discriminate between the two explanations, and whether indeed the peculiar active/passive quiescent state phenomenology of this system is directly linked to the jet itself.  Regardless of the correct interpretation, the data presented here, in combination with recent radio and sub-mm results from higher luminosity systems, demonstrate that jets from black hole X-ray binaries exhibit a high level of variability -- either in flux density or intrinsic spectral shape, or both -- across a wide spectrum of Eddington ratios. At first glance, this may be surprising, in the sense that one might expect that, as the accretion rate decreases, so does the total jet power and, as a consequence, its degree of variability. 
However, precisely as a result of a reduced jet power, the jet synchrotron emission becomes less absorbed, and the inner parts of the jet emit lower frequency radiation that they would at higher accretion rates/jet powers. In other words, at any given frequency, the SED of a quiescent/low jet power source is going to be dominated by emission occurring at smaller (as in, closer to the jet base) physical scales than for brighter X-ray states. Thus, in spite or the reduced overall flux levels, fractional variability can be safely expected to remain strong in the less powerful sources. 

\section*{Acknowledgements}
This paper makes use of the following ALMA data: ADS/JAO.ALMA\#2016.1.00773.S. ALMA is a partnership of ESO (representing its member states), NSF (USA) and NINS (Japan), together with NRC (Canada), MOST and ASIAA (Taiwan), and KASI (Republic of Korea), in cooperation with the Republic of Chile. The Joint ALMA Observatory is operated by ESO, AUI/NRAO and NAOJ.
The National Radio Astronomy Observatory is a facility of the National Science Foundation (NSF) operated under cooperative agreement by Associated Universities, Inc.
JCAM-J is the recipient of an Australian Research Council Future Fellowship (FT140101082) funded by the Australian government. EG is grateful to Julien Malzac for his insightful comments on an earlier version of this manuscript.




\bibliographystyle{mnras}

\bsp	
\label{lastpage}
\end{document}